\documentclass{desyproc}

\begin{document}
\title{Flavour Theory and the LHC Era}

\author{{\slshape Andrzej J.~Buras}\\[1ex]
Physik-Department, Technische Universit\"at M\"unchen,
D-85748 Garching, Germany\\
TUM Institute of Advanced Study, Lichtenbergstr. 2a, D-85748 Garching, Germany}

\contribID{xy}  
\confID{1964}
\desyproc{DESY-PROC-2010-01}
\acronym{PLHC2010}

\maketitle

\begin{abstract}
 This decade should make
 a significant progress towards the Theory of Flavour and the main goal of this
 talk is to transfer this believe to my colleagues in the
  particle physics community. Indeed a significant part of this decade 
could turn out to be the Flavour Era with participation of the LHC, Belle II, 
Super-Flavour-Facility and dedicated Kaon and lepton flavour violation 
experiments. Selected superstars of flavour physics listed below will play 
a prominent role in these events. In this writeup the leading role is played 
by the {\it prima donna} of 2010: CP violation in $B_s$ system.
\end{abstract}

\section{Introduction}
\begin{figure}[hb]
\centerline{\includegraphics[width=0.65\textwidth]{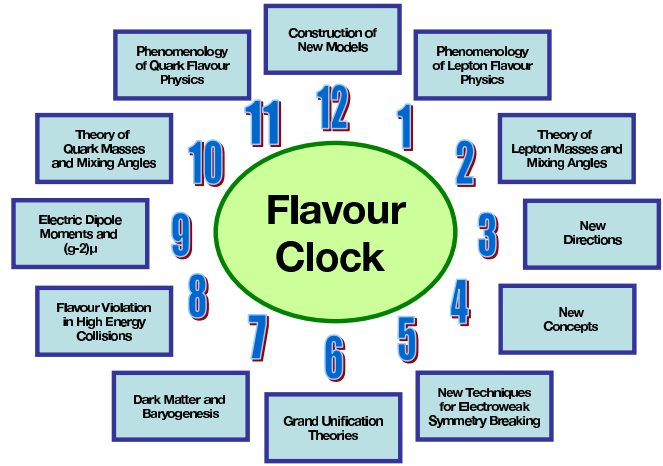}}
\caption{Working towards the Theory of Flavour around the Flavour Clock.}\label{Fig:1}
\end{figure}

In our search for a fundamental theory of elementary particles we need to
improve our understanding of flavour \cite{Buras:2009if,Isidori:2010kg}. This is clearly a very ambitious goal that
requires the advances in different directions as well as continuos efforts of
many experts day and night, as depicted with the help of a ''Flavour Clock''
in Figure~\ref{Fig:1}.

Despite  the impressive success of the CKM picture of flavour changing 
interactions \cite{Cabibbo:1963yz} in which also the GIM mechanism \cite{Glashow:1970gm} for the
suppression of flavour changing neutral currents (FCNC) 
plays a very important role, there are many open questions of 
theoretical and experimental nature that should be answered before we
can claim to have a theory of flavour.
Among the basic questions in flavour physics that could be answered in the 
present decade are the following ones:
\begin{enumerate}
\item
What is the fundamental dynamics behind the electroweak symmetry breaking 
that very likely plays also an important role in flavour physics?
\item
Are there any new flavour symmetries that could
help us to understand the existing hierarchies of fermion masses and the 
hierarchies in the quark and lepton flavour violating interactions?
\item
Are there any flavour violating interactions that are not governed by 
the SM Yukawa couplings? In other words, is  Minimal Flavour Violation 
(MFV)
the whole story?
\item
Are there any additional {\it flavour violating} CP-violating (CPV) phases that could 
explain certain anomalies present in the flavour data and 
simultaneously play
a role in the explanation of the observed baryon-antibaryon asymmetry 
in the universe (BAU)?
\item
Are there any {\it flavour conserving} CPV phases that could also help
in explaining the flavour anomalies in question and would be signalled 
in this decade  through  enhanced electric dipole moments (EDMs) of the
neutron, the electron and of other particles?
\item
Are there any new sequential heavy quarks and leptons of the 4th 
generation and/or new fermions with exotic quantum numbers like 
vectorial fermions?
\item
Are there any elementary neutral and charged scalar particles  
with masses below 1~TeV and having a significant impact on flavour physics?
\item
Are there any new heavy gauge bosons representing an enlarged gauge 
symmetry group?
\item
Are there any relevant right-handed (RH) weak currents that would help us to
make our fundamental theory parity conserving at short distance scales 
well below those explored by the LHC?
\item
How would one successfully address all these question if the breakdown of 
the electroweak symmetry would turn out to be of a non-perturbative origin? 
\end{enumerate}

An important question is the following one:
will some of these questions be answered through the interplay of high
energy processes explored by the LHC with low energy precision experiments 
or are the relevant scales of fundamental flavour well beyond the energies
explored by the LHC and future colliders in this century? The existing 
tensions in some of the corners of the SM and still a rather big room for
new physics (NP) contributions in rare decays of mesons and leptons and 
CP-violating observables including in particular EDMs give us hopes that indeed several phenomena required to answer at least
some of these questions could be discovered in this decade.

\section{Superstars of Flavour Physics in 2010-2015}
In this decade we will be able to resolve the short distance scales by more
than an order of magnitude, extending the picture of fundamental physics 
down to scales $5\cdot 10^{-20}$m with the help of the LHC. Further resolution 
down to scales as short as $10^{-21}$m or even shorter scales
 should be possible with the help of 
high precision experiments in which flavour violating processes will play a 
prominent role.

As far as high precision experiments are concerned a number of selected 
processes and observables will in my opinion play the leading role in 
learning about the NP in this new territory. This selection is based on 
the sensitivity to NP and theoretical cleanness. The former can be increased 
with the increased precision of experiments and the latter can improve with
the progress in theoretical calculations, in particular the non-perturbative
ones like the lattice simulations.

My superstars for the coming years are as follows:
\begin{itemize}
\item
The mixing induced CP-asymmetry $S_{\psi\phi}(B_s)$ that is
 tiny in the SM: $S_{\psi\phi}\approx 0.04$. The asymmetry
 $S_{\phi\phi}(B_s)$ is also important. It is also
 very strongly suppressed 
in the SM and is sensitive to NP similar to the one explored through 
the departure of $S_{\phi K_S}(B_d)$ from $S_{\psi K_S}(B_d)$ 
\cite{Fleischer:2007wg}.
\item
The rare decays $B_{s,d}\to\mu^+\mu^-$ that could be enhanced in certain 
NP scenarios by an order of magnitude with respect to the SM values.
\item
The angle $\gamma$ of the unitarity triangle (UT) that can be precisely 
measured
through tree level decays.
\item
$B^+\to\tau^+\nu_\tau$ that is sensitive to charged Higgs particles.
\item
The rare decays $K^+\to\pi^+\nu\bar\nu$ and $K_L\to\pi^0\nu\bar\nu$ that
belong to the theoretically cleanest decays in flavour physics.
\item
Numerous angular symmetries and asymmetries in $B\to K^*l^-l^-$.
\item
Lepton flavour violating decays like $\mu\to e\gamma$, $\tau\to e\gamma$, 
$\tau\to\mu\gamma$, decays with three leptons in the final state and 
$\mu-e$ conversion in nuclei.
\item
Electric dipole moments of the neutron, the electron, atoms and leptons.
\item
Anomalous magnetic moment of the muon $(g-2)_\mu$ that indeed seems to
be ''anomalous'' within the SM even after the inclusion of radiative corrections.
\item
The ratio $\varepsilon'/\varepsilon$ in $K_L\to\pi\pi$ decays 
which is known experimentally within 
$10\%$ and which should gain in importance in this decade due to improved 
lattice calculations.
\end{itemize}

Clearly, there are other stars in flavour physics but I believe that the 
ones above will play the crucial role in our search for the theory of 
flavour. Having experimental results on these decays and observables with
sufficient precision accompanied by improved theoretical calculations will
exclude several presently studied models reducing thereby our exploration
of short distance scales to a few avenues.

In the rest of this presentation I will discuss some of these decays in 
the context of the basic questions in flavour physics listed previously.
In particular we will collect a number of messages on NP which result from 
the recent and not so recent model independent studies and detailed analyses
of concrete numerous beyond the SM models (BSM). In this context the role of correlations 
between various observables implying various patterns of flavour violation 
characteristic for various concrete models should be strongly emphasized. 
Recent reviews can be found in \cite{Buras:2009if,Isidori:2010kg}. In the 
context of $B_{s,d}$-mixing and related NP see a very detailed recent 
analysis in  \cite{Lenz:2010gu}.

\section{Beyond the Standard Model (BSM)}
During the last 35 years several extensions of the SM have been proposed and
analyzed in the rich literature. In particular in the last 10 years, after 
the data on $B_{d,s}$ decays, $B^0_{d,s}-\bar B^0_{d,s}$ mixing and 
related CP violation improved considerably and the bounds on lepton flavour 
violating decays became stronger, useful model independent 
analyses of FCNC processes could be performed. Moreover several 
extensive analyses of 
the full spectrum of flavour violating processes
in the context of specific BSM scenarios have been published.
\subsection{Minimal Flavour Violation}
Among the model independent approaches in flavour physics the most prominent 
role is played by MFV \cite{D'Ambrosio:2002ex,Buras:2000dm} in which flavour violation including CP violation 
originates entirely from the SM Yukawa couplings. This approach naturally 
suppresses FCNC processes to the level observed experimentally even in the 
presence of new particles with masses of a few hundreds GeV. It also implies 
specific correlations between various observables, which are most stringent 
in the so-called constrained MFV (CMFV) \cite{Buras:2000dm}
 in which only the SM operators are
assumed to be relevant. Basically MFV reduces to CMFV when only one Higgs 
doublet is present.

A particularly interesting set-up is obtained introducing flavour-blind 
CPV phases compatible with the MFV symmetry principle~\cite{Kagan:2009bn,Colangelo:2008qp,Mercolli:2009ns,Paradisi:2009ey,Ellis:2007kb}.

As recently shown in~\cite{Buras:2010mh}, the general formulation of the
 MFV hypothesis with flavour-blind CPV phases (FBPh) applied to a general
 two Higgs doublet
 model  is very effective in
suppressing FCNCs to a level consistent
 with experiments,
leaving open the possibility of sizable non-standard effects also in 
CPV observables. In what follows we will call this model 
${\rm 2HDM_{\overline{MFV}}}$ with the ''bar'' on MFV indicating the 
presence of FBPhs.
As discussed in~\cite{Buras:2010mh}, the ${\rm 2HDM_{\overline{MFV}}}$
can accommodate a large CP-violating phase in $B_s$ mixing, as hinted by CDF and
D0 data~\cite{Aaltonen:2007he,Abazov:2010hv,Aaltonen:2010xx}, while ameliorating
simultaneously the observed anomaly in the relation between $\epsilon_K$ and 
$S_{\psi K_S}$~\cite{Lunghi:2008aa,Buras:2008nn}.

On general grounds, it is natural to expect that FBPhs
 contribute
also to CPV flavour-conserving processes, such as the EDMs. Indeed, the choice
adopted in~\cite{D'Ambrosio:2002ex} to assume the Yukawa couplings as the unique
breaking terms of both the flavour symmetry and the CP symmetry, was motivated
by possibly too large effects in EDMs with generic FBPhs.
This potential problem has indeed been confirmed by the recent model-independent
analysis in~\cite{Batell:2010qw}.

In \cite{Buras:2010zm} the correlations
between EDMs and CP
violation in $B_{s,d}$ mixing in ${\rm 2HDM_{\overline{MFV}}}$ including
 FBPhs in Yukawa interactions and  the Higgs potential have been studied 
in detail.
It has been shown that in both cases the upper bounds on 
EDMs of the neutron and the atoms 
do not forbid sizable non-standard CPV effects in $B_{s}$ mixing.
However, if a large CPV phase in $B_s$ mixing will be confirmed, this
will imply hadronic EDMs very close to their present experimental bounds,
within the reach of the next generation of experiments, as well as
$ Br(B_{s,d}\to\mu^+\mu^-)$ typically largely enhanced over its
SM expectation. The two flavour-blind CPV mechanisms can be distinguished
through the correlation between $S_{\psi K_S}$ and $S_{\psi\phi}$ that is
strikingly different if only one of them is relevant. Which of these two
CPV mechanisms dominates depends on the precise values of $S_{\psi\phi}$
and $S_{\psi K_S}$, as well as on the CKM phase (as determined by tree-level
processes). Current data seems to show a mild preference for a {\it hybrid}
scenario where both these mechanisms are at work. I will be a bit more 
explicit about this result below.
\subsection{Beyond Minimal Flavour Violation}
There is a number of explicit BSM models that introduce new sources of 
flavour violation and CP violation beyond those present in the MFV 
framework discussed above. Among them the Littlest Higgs Model with 
T-parity (LHT), the Randall-Sundrum model without and with custodial protection (RSc), 
various supersymmetric flavour models, $Z^\prime$-models, models with 
vectorial new quarks, the SM extended by the fourth sequential generation of quarks and leptons (SM4) and multi-Higgs doublet models are the ones in 
which most extensive flavour analyses have been performed. Most of them 
have  been 
reviewed in some details in \cite{Buras:2009if}, where the relevant references can be found. 
I will concentrate  in this presentation on very recent developments and will
only recall some of the most interesting results of these older analyses if
necessary.

During the  second half of 2009 and also in 2010 the flavour analyses in the 
framework of 
the 2HDM with and without MFV and also the SM4 became popular. 
The ${\rm 2HDM_{\overline{MFV}}}$ has been already briefly discussed above. The SM4 introduces three
new mixing angles $s_{14}$, $s_{24}$, $s_{34}$ and two new phases in the 
quark sector and can still have a significant impact on flavour 
phenomenology. Most recent extensive analyses of FCNC processes in the SM4
can be found in \cite{Hou:2005yb,Soni:2008bc,Bobrowski:2009ng,Buras:2010pi}.
 More about it later.

Next, let me mention an effective theory approach in which
the impact of RH currents in both charged- and neutral-current 
flavour-violating processes has been  analysed \cite{Buras:2010pz}. 
While RH currents 
are present in several supersymmetric flavour models, in RS models and of 
course in left-right symmetric models based on the gauge group
$SU(2)_L \times SU(2)_R \times U(1)_{B-L}$ (see \cite{Maiezza:2010ic,Guadagnoli:2010sd} for most recent papers),
the recent phenomenological interest in these models 
 originated in tensions between inclusive and exclusive
 determinations of the elements of the CKM matrix $|V_{ub}|$ and  $|V_{cb}|$. 
 It could be that these tensions are due to the underestimate of theoretical 
 and/or experimental uncertainties. Yet, it is a fact, as pointed out
 and analyzed recently in particular in 
\cite{Crivellin:2009sd,Chen:2008se,Feger:2010qc}, 
 that the presence of RH  currents could either remove 
 or significantly weaken some of these tensions, especially in the 
 case of $|V_{ub}|$. Implications of this setup
for other observables, in particular 
 FCNC processes
without specifying
 the fundamental theory in detail but only assuming its global symmetry
 and the pattern of its breakdown have been analyzed in \cite{Buras:2010pz}. 
As we will see this approach can be considered as a minimal flavour 
violating scenario in the RH sector and will be called RHMFV in what 
follows.
I will return to the results of this work below. 

Finally, recent studies of flavour violating processes in models for fermion 
masses and mixings  \cite{Altmannshofer:2009ne,Lalak:2010bk,Dudas:2010yh}, indicate that a full theory 
of flavour has to involve at a certain level non-MFV interactions.

\section{Waiting for Signals of NP in FCNC Processes}
\subsection{General Remarks}
The last decade has established that flavour-changing and CPV processes in
$B_{s,d}$ and K systems are on the whole well described by the SM. The same 
applies to electroweak precision tests. This implies automatically tight 
constraints on flavour-changing phenomena beyond the SM and a potential 
problem for a natural solution of the hierarchy problem and other problems 
listed in the Introduction, several of which require the presence of NP not
far from the electroweak scale.

It is evident from various model-independent studies that NP at the TeV scale
must have a non-generic flavour structure in order to satisfy existing 
constraints. Moreover, in order to avoid fine tuning of parameters, natural
protection mechanisms suppressing FCNCs generated by NP are required. In 
addition to MFV and GIM, RS-GIM, T-parity in Littlest Higgs models, alignment
and degeneracy, most familiar from supersymmetric models and generally flavour 
symmetries (abelian and non-abelian) have been invented for this purpose.
Last but certainly not least, custodial symmetries, like the ones related to
the Higgs system and relevant for electroweak precision tests, can be used to
 suppress specific flavour-violating neutral gauge boson couplings.

It should be emphasized that only protection mechanisms that are stable 
under radiative corrections can be considered as solutions to flavour 
problems and considerations of protection mechanisms only at tree level are
insufficient. 
In this context let us recall that the standard assignment of the $SU(2)_L\times U(1)_Y$ quark charges, identified long ago by Glashow, Iliopoulos, and Maiani (GIM)~\cite{Glashow:1970gm},
forbids tree-level flavour-changing couplings of the quarks to the 
SM neutral gauge bosons. This mechanism is only violated
at the loop level and the FCNC processes are strongly suppressed by the 
products of CKM elements and mass splittings of quarks or leptons carrying 
the same electric charge. Only in processes involving the top quark exchanges 
is GIM strongly broken but in a calculable manner and the pattern of this 
breakdown seems to agree with experiment although the tests of this pattern
have to be still very much improved.
 
In the case of only one Higgs doublet, 
namely within the SM, this structure is effective also 
in eliminating  possible dimension-four  
FCNC couplings of the quarks to the Higgs field. 
While the $SU(2)_L\times U(1)_Y$ assignment of quarks and leptons 
can be considered as being well established, much less is known 
about the Higgs sector of the theory. In the presence 
of more than one Higgs field the appearance of tree-level FCNC 
is not automatically forbidden by the standard assignment 
of the $SU(2)_L\times U(1)_Y$ fermion charges: additional conditions have to 
be imposed on the model in order to guarantee a sufficient suppression of 
FCNC processes~\cite{Glashow:1976nt,Paschos:1976ay}. 
The absence of renormalizable couplings contributing 
at the tree level to FCNC processes, in multi-Higgs 
models, goes under the name of Natural Flavour Conservation (NFC)
hypothesis.

It has been pointed out recently \cite{Buras:2010mh} that the MFV hypothesis 
is more stable 
in suppressing FCNCs than the hypothesis of NFC alone
when quantum corrections are taken into account. Indeed the NFC 
hypothesis is usually based on a $U(1)_{PQ}$ symmetry that has to be 
broken in order to avoid massless scalars. NFC can also be enforced by a 
 $Z_2$ symmetry. However, it turns out that also this symmetry is
 insufficient to protect FCNCs when radiative corrections
are considered. On the other hand MFV hypothesis based on continuous flavour 
symmetries  is more powerful. Thus 30 years after the seminal papers of 
Glashow, Weinberg and Paschos, the hypothesis of NFC can be replaced by the more 
powerful and more general hypothesis of MFV. Other recent interesting 
analyzes of 2HDMs can be found in 
\cite{Botella:2009pq,Pich1,Dobrescu:2010rh,Braeuninger:2010td}.

\subsection{Three Strategies in Waiting for NP in Flavour Physics}
Particle physicists are waiting eagerly for a solid evidence of NP for the
last 30 years. Except for neutrino masses, the BAU and dark matter, no clear 
signal emerged so far.
 While waiting several strategies for finding NP have been 
developed. They can be divided roughly into three classes.
\subsubsection{Precision calculations within the SM}
Here basically the goal is to calculate the background to NP coming from
the known dynamics of the SM. At first sight this approach is not very
exciting. Yet, in particular in flavour physics, where the signals of 
NP are generally indirect, this approach is very important. From my 
point of view, being involved more than one decade in calculations of 
higher order QCD corrections \cite{Buras:1998raa}, 
I would claim that for most interesting
decays these perturbative and renormalization group improved calculations
reached already the desired level. The most advanced NNLO QCD calculations 
have been done for $B\to X_s\gamma$, $K^+\to\pi^+\nu\bar\nu$, $B\to X_s l^+l^-$ 
and recently for $\varepsilon_K$  \cite{Brod:2010mj}. See also the two loop
electroweak contributions to $K\to\pi\nu\bar\nu$ \cite{Brod:2010hi}.

The main progress is now required from lattice groups. Here the main goals 
for the coming years are more accurate values of weak decay constants 
$F_{B_{d,s}}$ and various $\hat B_i$ parameters relevant for $B_{d,s}$ physics.
For $K^0-\bar K^0$ mixing the relevant parameter $\hat B_K$ is now known 
with an accuracy of $4\%$ \cite{Antonio:2007pb}. An impressive achievement. Let us hope that 
also the parameters $B_6$ and $B_8$, relevant for 
$\varepsilon^\prime/\varepsilon$ will be known with a similar accuracy 
within this decade.

Clearly further improvements on the hadronic part of two-body 
non-leptonic decays is mandatory in order to understand more precisely 
the direct CP violation in $B_{s,d}$ decays. 
\subsubsection{The Bottom-Up Approach}
In this approach one constructs effective field theories involving 
only light degrees 
of freedom including the top quark in which the structure of the effective 
Lagrangians is governed by the symmetries of the SM and often other 
hypothetical symmetries. This approach is rather powerful in the case of
electroweak precision 
studies and definitely teaches us something about $\Delta F=2$ 
transitions. However, except for the case of  MFV and closely related 
approaches based on flavour symmetries, the bottom-up approach ceases, 
in my view, to be useful in $\Delta F=1$ decays, 
because of very many operators that are allowed to appear
in the effective Lagrangians with coefficients that are basically 
unknown \cite{Grzadkowski:2010es}. In this 
approach then the correlations between various $\Delta F=2$ and $\Delta F=1$ 
observables in $K$, $D$, $B_d$ and $B_s$ systems are either not visible or 
very weak, again except MFV, CMFV or closely related approaches. Moreover 
the correlations between flavour violation in low energy processes and 
flavour violation in high energy processes to be studied soon at the LHC 
is lost. Again MFV belongs to a few exceptions.
\subsubsection{The Top-Down Approach}
My personal view shared by some of my colleagues is that the top-down 
approach is more useful in flavour physics. Here one constructs first 
a specific model with heavy degrees of freedom. For high energy processes,
where the energy scales are of the order of the masses of heavy particles 
one can directly use this ``full theory'' to calculate various processes 
in terms of the fundamental parameters of a given theory. For low energy 
processes one again constructs the low energy theory by integrating out 
heavy particles. The advantage over the previous approach is that now the 
coefficients of the resulting local operators are calculable in terms of 
the fundamental parameters of this theory. In this manner correlations between 
various observables belonging to different mesonic systems and correlations 
between low energy and high-energy observables are possible. Such correlations 
are less sensitive to the free parameters than separate observables and 
represent patterns of flavour violation characteristic for a given theory. 
These correlations can in some models differ strikingly from the ones of 
the SM and of the MFV approach.
\subsection{Anatomies of explicit models}
Having the last strategy in mind my group at the Technical University Munich, 
consisting dominantly of diploma students, PhD students and young post--docs 
investigated in the last decade flavour violating processes with the emphasis 
put on FCNC processes, in the following models: CMFV, MFV, MFV-MSSM, 
$Z^\prime$-models, general MSSM, a model with a universal flat 5th dimension, 
the Littlest Higgs model (LH), the Littlest Higgs model with T-parity (LHT),
SUSY-GUTs,  
Randall-Sundrum model with custodial protection (RSc), flavour blind MSSM 
(FBMSSM), three classes of supersymmetric flavour models with the dominance 
of left-handed currents ($\delta$LL model), the dominance of right-handed currents 
 (AC model) and   models 
with equal 
strength of left- and right-handed currents (RVV2 and AKM models), the last comments applying only
to the NP part. This year we have analyzed the SM4, the 
${\rm 2HDM_{\overline{MFV}}}$ and finally quark flavour mixing with RH 
currents in an effective theory approach RHMFV.
These analyses where dominated by quark flavour physics, but in the case of the
LHT, FBMSSM, supersymmetric flavour models and the  SM4 lepton flavour 
violation has also been studied in detail. 

As a partial review of this work appeared already in \cite{Buras:2009if}
 with various 
correlations presented in Figures 5 - 11 of that paper I will not discuss
them in detail here. In \cite{Buras:2009if} numerous references (301) to
our papers and studies by other 
groups can be found. The detailed discussion of the supersymmetric 
flavour models ($\delta$LL, AC, RVV2, AKM) can be found in \cite{Altmannshofer:2009ne}.

\newcommand{\three}{{\color{red}$\bigstar\bigstar\bigstar$}}
\newcommand{\two}{{\color{blue}$\bigstar\bigstar$}}
\newcommand{\one}{{\color{black}$\bigstar$}}

\begin{table}[t]
\addtolength{\arraycolsep}{4pt}
\renewcommand{\arraystretch}{1.5}
\centering
\begin{tabular}{|l|c|c|c|c|c|c|c|c|}
\hline
&  AC & RVV2 & AKM  & $\delta$LL & FBMSSM & LHT & RSc & 4G
\\
\hline\hline
$D^0-\bar D^0$& \three & \one & \one & \one & \one & \three & ? & \two
\\ \hline
$\epsilon_K$& \one & \three & \three & \one & \one & \two & \three & \two
\\ \hline
$ S_{\psi\phi}$ & \three & \three & \three & \one & \one & \three & \three & \three
\\ \hline\hline
$S_{\phi K_S}$ & \three & \two & \one & \three & \three & \one & ?  & \two
\\ \hline
$A_{\rm CP}\left(B\rightarrow X_s\gamma\right)$ & \one & \one & \one & \three & \three & \one & ? & \one
\\ \hline
$A_{7,8}(K^*\mu^+\mu^-)$ & \one & \one & \one & \three & \three & \two & ?
& \two
\\ \hline
$B_s\rightarrow\mu^+\mu^-$ & \three & \three & \three & \three & \three & \one & \one & \three
\\ \hline
$K^+\rightarrow\pi^+\nu\bar\nu$ & \one & \one & \one & \one & \one & \three & \three & \three
\\ \hline
$K_L\rightarrow\pi^0\nu\bar\nu$ & \one & \one & \one & \one & \one & \three & \three & \three
\\ \hline
$\mu\rightarrow e\gamma$& \three & \three & \three & \three & \three & \three & \three & \three
\\ \hline
$\tau\rightarrow \mu\gamma$ & \three & \three & \one & \three & \three  & \three & \three & \three
\\ \hline
$\mu + N\rightarrow e + N$& \three & \three & \three & \three & \three & \three & \three & \three
\\ \hline\hline
$d_n$& \three & \three & \three & \two & \three & \one & \three & \one
\\ \hline
$d_e$& \three & \three & \two & \one & \three & \one & \three & \one
\\ \hline
$\left(g-2\right)_\mu$& \three & \three & \two & \three & \three & \one & ?
& \one
\\ \hline

\end{tabular}
\renewcommand{\arraystretch}{1}
\caption{\small
``DNA'' of flavour physics effects for the most interesting observables in a selection of SUSY
and non-SUSY models. \three\ signals large NP effects, \two\ visible but small NP effects and \one\
implies that the given model does not predict sizable NP effects in that observable. From \cite{Altmannshofer:2009ne}.}
\label{tab:DNA}
\end{table}

The ``DNA'' of flavour physics effects for the most interesting observables 
constructed in \cite{Altmannshofer:2009ne} and extended by the recent results obtained in the 
SM4 is presented in Table~\ref{tab:DNA}. This table only indicates whether large, moderate 
or small NP effects in a given observable are still allowed in a given model 
but does not exhibit correlations between various observables characteristic 
for a given model. Such correlations can be found in \cite{Buras:2009if}
 and original papers quoted there.
I will summarize the most striking ones later on.
\boldmath
\subsection{$\varepsilon_K$-anomaly and related tensions}
\unboldmath
It has been pointed out in \cite{Buras:2008nn} that the SM prediction for $\varepsilon_K$ 
implied by the measured value of $S_{\psi K_S}=\sin 2\beta$, the ratio 
$\Delta M_d/\Delta M_s$ and the value of $|V_{cb}|$ turns out to be too
small to agree well with experiment. This tension between $\varepsilon_K$ and
 $S_{\psi K_S}$ has been pointed out from a different perspective in
\cite{Lunghi:2008aa}.
These findings have been confirmed by a 
UTfitters  analysis \cite{UTfit-web}. 
The CKMfitters having a different treatment of uncertainties find less significant effects \cite{Lenz:2010gu}.

 The main reasons 
for this tension are on the one hand a decreased value of the relevant non-perturbative 
parameter $\hat B_K=0.724\pm0.008\pm0.028$ \cite{Antonio:2007pb}
resulting from unquenched lattice 
calculations and on the other hand the decreased value of $\varepsilon_K$ in 
the SM arising from a multiplicative factor, estimated first to be 
$\kappa_\varepsilon=0.92\pm0.02$ \cite{Buras:2008nn}. This factor took into account the departure 
of $\phi_\varepsilon$ from $\pi/4$ and the long distance (LD) effects in 
${\rm Im}\Gamma_{12}$ in the $K^0-\bar K^0$ mixing. The recent inclusion of LD effects 
in ${\rm Im}M_{12}$ modified this estimate to $\kappa_\varepsilon=0.94\pm0.02$ 
\cite{Buras:2010pza}. 
Very recently also NNLO-QCD corrections to the QCD factor $\eta_{ct}$ in 
$\varepsilon_K$ \cite{Brod:2010mj} have 
been calculated enhancing the value of $\varepsilon_K$ by $3\%$. Thus while 
in \cite{Buras:2008nn} the value $|\varepsilon_K|_{\rm SM}=(1.78\pm0.25)\cdot 10^{-3}$ has 
been quoted and with the new estimate of LD effects and new input one 
finds $|\varepsilon_K|_{\rm SM}=(1.85\pm0.22)\cdot 10^{-3}$, including NNLO
corrections gives the new value 
\begin{equation}\label{epnew}
|\varepsilon_K|_{\rm SM}=(1.92\pm0.25)\cdot 10^{-3},
\end{equation}
significantly closer to the experimental value 
$|\varepsilon_K|_{\rm exp}=(2.23\pm0.01)\cdot 10^{-3}$. This result is 
compatible with \cite{Brod:2010mj,Lenz:2010gu} although the central value in (\ref{epnew}) 
is sensitive 
to the input parameters, in particular the value of $\sin 2\beta$.

Consequently, the $\varepsilon_K$-anomaly softened considerably but it is still 
alive. Indeed, the $\sin 2\beta=0.74\pm 0.02$ from UT fits is visibly larger
than the experimental value $S_{\psi K_S}=0.672\pm 0.023$. The difference is 
even larger if one wants to fit $\varepsilon_K$ exactly: 
$\sin 2\beta\approx 0.80$  \cite{Lunghi:2008aa,Buras:2008nn}.

One should also recall the tension between inclusive and exclusive determinations with the exclusive ones in the ballpark of $3.5\cdot 10^{-3}$ and the 
inclusive ones typically above $4.0\cdot 10^{-3}$.

As discussed in \cite{Lunghi:2008aa,Buras:2008nn} a small negative NP phase $\varphi_{B_d}$ in 
$B^0_d-\bar B^0_d$ mixing would solve some of these problems. Indeed we have 
then 
\begin{equation}
S_{\psi K_S}(B_d) = \sin(2\beta+2\varphi_{B_d})\,, \qquad
S_{\psi\phi}(B_s) =  \sin(2|\beta_s|-2\varphi_{B_s})\,,
\label{eq:basic}
\end{equation}
where the corresponding formula for $S_{\psi\phi}$ in the presence of 
a NP phase $\varphi_{B_s}$ in 
$B^0_s-\bar B^0_s$ mixing has also been given. With a negative $\varphi_{B_d}$ 
the true $\sin 2\beta$ is larger than $S_{\psi K_S}$, implying a higher value
on $|\varepsilon_K|$, in reasonable agreement with data and a better UT-fit. This 
solution would favour the inclusive value of $|V_{ub}|$.

Now with a  universality hypothesis of $\varphi_{B_s}=\varphi_{B_d}$ 
\cite{Ball:2006xx,Buras:2008nn},
a negative $\varphi_{B_d}$ would automatically imply an enhanced value of
$S_{\psi\phi}$ which in 
view of $|\beta_s|\approx 1^\circ$ amounts to roughly 0.04 in the SM. However, 
in order to be in agreement with the experimental value of $S_{\psi K_S}$ this 
type of NP would imply $S_{\psi\phi}\le 0.25$.

The  universality hypothesis of $\varphi_{B_s}=\varphi_{B_d}$ in 
\cite{Ball:2006xx,Buras:2008nn} was clearly
ad hoc. Recently, in view of the enhanced value of $S_{\psi\phi}$ at CDF and 
D0 a more dynamical origin of this relation has been discussed by other 
authors and different relations between these two phases corresponding still
to a different dynamics have been discussed in the literature. 
 Let us elaborate
on this topic in more detail.
\boldmath
\subsection{Facing an enhanced CPV in the $B_s$ mixing}
\unboldmath
Possibly the most important highlight in flavour physics in 2008, 2009 
\cite{Aaltonen:2007he} and even 
more in 2010 was the enhanced value of $S_{\psi\phi}$ measured by the CDF and 
D0 collaborations, seen either directly or indirectly through the 
correlations with various semi-leptonic asymmetries. While in 2009 and in 
the Spring of 2010  \cite{Abazov:2010hv}, the messages from Fermilab indicated good prospects 
for $S_{\psi\phi}$ above 0.5, the recent messages from ICHEP 2010 in Paris, 
softened such hopes significantly \cite{Aaltonen:2010xx}.
Both CDF and D0 find the enhancement 
by only one $\sigma$. Yet, this does not yet preclude $S_{\psi\phi}$ above 0.5, 
which would really be a fantastic signal of NP. But $S_{\psi\phi}$ below 
0.5 appears more likely at present. Still even a value of 0.2 would 
be exciting. Let us hope that the future data from Tevatron and in 
particular from the LHCb, will measure this asymmetry with sufficient 
precision so that we will know to which extent NP is at work here. 
One should also hope that the large CPV in dimuon CP asymmetry from D0,
 that triggered 
new activities, will be better understood. I have nothing to add here
at present and can only refer to numerous papers  
\cite{Dobrescu:2010rh,Ligeti:2010ia,Blum:2010mj,Lenz:2010gu,Bauer:2010dg}.

Leaving the possibility of $S_{\psi\phi}\ge 0.5$ still open but keeping in 
mind that also $S_{\psi\phi}\le 0.25$ could turn out to be the final value, 
let us investigate how different models would face these two different 
results and what kind of dynamics would be behind these two scenarios.
\boldmath
\subsubsection{$S_{\psi\phi}\ge 0.5$}
\unboldmath
Such large values can be obtained in the RSc model due to KK gluon 
exchanges and also heavy neutral KK electroweak gauge boson exchanges. 
In the  supersymmetric flavour model with the dominance of right-handed 
currents  like the AC model, double Higgs penguins constitute the dominant NP 
contributions responsible for $S_{\psi\phi}\ge 0.5$, while in the RVV2 model where NP left-handed current 
contributions are equally important, also gluino boxes are relevant.
On the operator level, it is LR {\it scalar} operator
 which is primarly responsible 
for this enhancement.

Interestingly the SM4 having only $(V-A)*(V-A)$ operators is also capable 
in obtaining 
high values of $S_{\psi\phi}$ \cite{Hou:2005yb,Soni:2008bc,Buras:2010pi} but not as easily as the RSc, AC and RVV2 models.
The lower scales of NP in the SM4 relative to the latter models and the
non-decoupling effects of $t^\prime$ compensate to some extent the absene 
of LR scalar operators. In the LHT model where only $(V-A)*(V-A)$ operators 
are present 
and the NP enters at higher scales than in the SM4, $S_{\psi\phi}$ above 
0.5 is out of reach \cite{Blanke:2009am}.

All these models contain new sources of flavour  and CP violation 
and it is not surprising that in view of many parameters involved large 
values of $S_{\psi\phi}$ can be obtained. The question then arises whether 
strongly
enhanced values of this asymmetry would uniquely imply new sources of 
flavour violation beyond the MFV hypothesis. The answer to this question is as 
follows:
\begin{itemize}
\item
In models with MFV and FBPhs set to zero, $S_{\psi\phi}$
remains indeed SM-like.
\item
In supersymmetric models with MFV and non-vanishing FBPhs and in the  FBMSSM, at both 
small and 
large $\tan\beta$, the supersymmetry constraints do not allow values 
of $S_{\psi\phi}$ visibly different from the SM value 
\cite{Altmannshofer:2009ne,Blum:2010mj}
\item
In the ${\rm 2HDM_{\overline{MFV}}}$ in which at one-loop both Higgs
doublets couple to up- and down-quarks, the interplay  of FBPh with 
the CKM matrix allows to obtain $S_{\psi\phi}\ge 0.5$ while satisfying 
all existing constraints \cite{Buras:2010mh}.
\end{itemize}

In the presence of a large $S_{\psi\phi}$
the latter model allows also for a simple 
and unique softening of the $\varepsilon_K$-anomaly and of the tensions in the 
UT analysis if the FBPh in the Yukawa interactions are the dominant source 
of new CPV. In this case the NP phases 
 $\varphi_{B_s}$ and $\varphi_{B_d}$ are related through 
\begin{equation}\label{BCGI}
 \varphi_{B_d}\approx\frac{m_d}{m_s}\varphi_{B_s}\approx \frac{1}{17} \varphi_{B_s},
\end{equation}
in visible contrast to the hypothesis $\varphi_{B_s}=\varphi_{B_d}$ of 
\cite{Ball:2006xx,Buras:2008nn}. 
Thus in this scenario large $\varphi_{B_s}$ required to obtain values of 
$S_{\psi\phi}$ above 0.5 imply a unique small shift in $S_{\psi K_S}$ 
that allows to lower $S_{\psi K_S}$ from 0.74 down to 0.70, 
that is closer to the experimental value $0.672\pm0.023$. This in turn 
implies that it is $\sin 2\beta=0.74$ and not  $S_{\psi K_S}=0.67$ that 
should be used in calculating $\varepsilon_K$ resulting in a value of 
 $\varepsilon_K\approx 2.0\cdot 10^{-3}$ within one $\sigma$ from the 
experimental value. The direct Higgs contribution to $\varepsilon_K$ 
is negligible because of small masses $m_{d,s}$. We should emphasize that 
once  $\varphi_{B_s}$ is determined from the data on $S_{\psi\phi}$ by means 
of (\ref{eq:basic}), the implications for $\varepsilon_K$ and  $S_{\psi K_S}$ are 
unique. It is remarkable that such a simple set up 
allows basically to solve all these tensions provided $S_{\psi\phi}$ is 
sufficiently above 0.5. The plots of $\varepsilon_K$ and  $S_{\psi K_S}$ 
versus $S_{\psi\phi}$ in \cite{Buras:2010mh} show this very transparently.
\boldmath
\subsubsection{$S_{\psi\phi}\approx 0.25$}
\unboldmath
Yet, as signalled recently by  CDF and D0 data \cite{Aaltonen:2010xx},  $S_{\psi\phi}$ could be 
smaller. In this case all non-MFV models listed above can reproduce 
such values and in particular this time also the LHT model \cite{Blanke:2009am} and another 
supersymmetric flavour model (AKM) analysed by us stay alive \cite{Altmannshofer:2009ne}.

Again MSSM-MFV cannot reproduce such values. On the other hand the
${\rm 2HDM_{\overline{MFV}}}$ can still provide interesting results. Yet 
as evident from the plots in \cite{Buras:2010mh} the FBPh in Yukawa interactions
cannot now solve the UT tensions. Indeed the relation in (\ref{BCGI}) 
precludes now any interesting effects in $\varepsilon_K$ and $S_{\psi K_S}$:
$S_{\psi\phi}$ and the NP phase  $\varphi_{B_s}$ are simply too small. 
Evidently, this time the relation
\begin{equation}\label{BG}
 \varphi_{B_d}=\varphi_{B_s}
\end{equation}
would be more appropriate.

Now, the analyses in \cite{Ligeti:2010ia,Blum:2010mj} indicate how such a relation could be 
obtained within the ${\rm 2HDM_{\overline{MFV}}}$. This time the FBPh in the 
Higgs potential are at work, the relation in (\ref{BG}) follows and 
the plots of $\varepsilon_K$ and  $S_{\psi K_S}$
versus $S_{\psi\phi}$ are strikingly modified: the dependence is much 
stronger and even moderate values of $S_{\psi\phi}$ can solve all tensions.
This time not  scalar LR operators but  scalar LL operators are responsible 
for this behaviour.

Presently it is not clear which relation between $\varphi_{B_s}$ and 
$\varphi_{B_d}$ fits best the data but the model independent analysis 
of \cite{Ligeti:2010ia} indicates that $\varphi_{B_s}$ should be significantly larger than 
$\varphi_{B_d}$, but this hierarchy appears to be smaller than in (\ref{BCGI}).
Therefore as pointed out in \cite{Buras:2010zm} in the ${\rm 2HDM_{\overline{MFV}}}$ the best agreement 
with the data is obtained by having these phases both in Yukawa interactions 
and the Higgs potential, which is to be expected  in any case.
Which of the two flavour-blind CPV mechanisms dominates depends on the value of
$S_{\psi\phi}$, which is still affected by a sizable experimental error, and
 also by the precise amount of NP allowed in $S_{\psi K_S}$.

Let us summarize the dynamical picture behind an enhanced value of 
$S_{\psi\phi}$ within ${\rm 2HDM_{\overline{MFV}}}$. For $S_{\phi\phi}\ge 0.7$ the 
FBPh in Yukawa interactions are expected to dominate. On the other hand 
for $S_{\phi\phi}\le 0.25$ the FBPh in the Higgs potential are expected to
dominate the scene. If $S_{\psi\phi}$ will eventually be found somewhere between 
0.3 and 0.6, a hybrid scenario analyzed in \cite{Buras:2010zm} would be most efficient 
although not as predictive as the cases in which only one of these 
two mechanism is at
work.
\boldmath
\subsection{Implications of an enhanced  $S_{\psi\phi}$}
\unboldmath
\subsubsection{Preliminaries}
Let us then assume that indeed $S_{\psi\phi}$ will be found to be significantly 
enhanced over the SM value. The studies of different observables in different 
models allow then immediately to make some concrete predictions on a number
of observables which makes it possible to distinguish different models. This 
is important as  $S_{\psi\phi}$ alone is insufficient for this purpose.

In view of space limitations I will discuss here only the implications for
$B_{s,d}\to \mu^+\mu^-$ and $K\to\pi\nu\bar\nu$ decays, which we declared to be
the superstars of the coming years. Subsequently I will make brief comments 
on a number of other superstars: EDMs, $(g-2)_\mu$, lepton flavour violation 
and $\varepsilon'/\varepsilon$.
\boldmath
\subsubsection{$S_{\psi\phi}\ge 0.5$ Scenario}
\unboldmath
The detailed studies of several models in which such high values of 
$S_{\psi\phi}$ can be attained imply the following pattern:
\begin{itemize}
\item
In the AC model and the ${\rm 2HDM_{\overline{MFV}}}$, $Br(B_{s,d}\to \mu^+\mu^-)$ will be 
automatically enhanced up to the present upper limit of roughly 
$3\cdot 10^{-8}$ from CDF and D0. The double Higgs penguins are responsible 
for this correlation 
\cite{Buras:2010mh,Buras:2010zm,Altmannshofer:2009ne}.
\item
In the SM4 this enhancement will be more moderate: up to $(6-9)\cdot 10^{-9}$, 
that is a factor of 2-3 above the SM value \cite{Soni:2008bc,Buras:2010pi}.
\item
In the non-abelian supersymmetric flavour model RVV2, 
$Br(B_{s,d}\to \mu^+\mu^-)$ can be enhanced up to a few $10^{-8}$ but it is not
uniquely implied due to the pollution of double-Higgs contributions through 
gluino boxes, that disturbs the correlation present in the AC model 
\cite{Altmannshofer:2009ne}.
\item
In the RSc, $Br(B_{s,d}\to \mu^+\mu^-)$ is SM-like independently of the value of 
 $S_{\psi\phi}$ \cite{Blanke:2008yr}. If the custodial protection for $Z$ flavour violating 
couplings is removed values of $10^{-8}$ are possible \cite{Blanke:2008yr,Bauer:2009cf}.
\end{itemize}

The question then arises what kind of implications does one have for 
$Br(B_{d}\to \mu^+\mu^-)$. Our studies show that
\begin{itemize}
\item
The ${\rm 2HDM_{\overline{MFV}}}$ implies automatically an enhancement of $Br(B_{d}\to \mu^+\mu^-)$ 
with the ratio of these two branching ratios governed solely 
by $|V_{td}/V_{ts}|^2$ and weak decay constants.
\item
This familiar MFV relation between the two branching ratios $Br(B_{s,d}\to \mu^+\mu^-)$ is strongly violated in non-MFV scenarios like AC and RVV2 models 
and as seen in Fig. 5 of \cite{Buras:2009if} 
taken from \cite{Altmannshofer:2009ne} for a given $Br(B_{s}\to \mu^+\mu^-)$ 
the range for $Br(B_{d}\to \mu^+\mu^-)$ can be large with the values of the
latter branching ratios being as high as $5\cdot 10^{-10}$.
\item
Interestingly, in the SM4, large $S_{\psi\phi}$ accompanied by 
large $Br(B_{s}\to \mu^+\mu^-)$  precludes a large departure of 
$Br(B_{d}\to \mu^+\mu^-)$ from the SM value $1\cdot 10^{-10}$ 
\cite{Buras:2010pi}.
\end{itemize}

We observe that simultaneous consideration of $S_{\psi\phi}$ and 
 $Br(B_{s,d}\to \mu^+\mu^-)$ can already help us in eliminating
some NP scenarios. Even more insight will be gained when 
$Br(K^+\to\pi^+\nu\bar\nu)$ and $Br(K_L\to\pi^0\nu\bar\nu)$ will be 
measured:
\begin{itemize}
\item
First of all the supersymmetric flavour models mentioned above predict by 
construction tiny NP contributions to $K\to\pi\nu\bar\nu$ decays. 
This is also the case of the ${\rm 2HDM_{\overline{MFV}}}$.
\item
In the RSc model significant enhancements of both branching ratios are
generally possible \cite{Blanke:2008yr,Bauer:2009cf}
 but not if $S_{\psi\phi}$ is large. Similar comments would
apply to the LHT model where the NP effects in $K\to\pi\nu\bar\nu$ can 
be larger than in the RSc \cite{Blanke:2009am}. However, the LHT model has difficulties to
reproduce a very large $S_{\psi\phi}$ and does not belong to this scenario.
\item
Interestingly, in the SM4 large $S_{\psi\phi}$, 
$Br(K^+\to\pi^+\nu\bar\nu)$ and $Br(K_L\to\pi^0\nu\bar\nu)$ can coexist 
with each other \cite{Buras:2010pi}.
\end{itemize}
\boldmath
\subsubsection{$S_{\psi\phi}\approx 0.25$ Scenario}
\unboldmath
In this scenario many effects found in the large $S_{\psi\phi}$ scenario
 are significantly weakend. Prominent exceptions are 
\begin{itemize}
\item
In the SM4, $Br(B_{s}\to \mu^+\mu^-)$ is not longer enhanced and can even 
be suppressed, while $Br(B_{d}\to \mu^+\mu^-)$ can be significantly enhanced
\cite{Buras:2010pi}.
\item
The branching ratios $Br(K^+\to\pi^+\nu\bar\nu)$ and $Br(K_L\to\pi^0\nu\bar\nu)$
can now be strongly enhanced in the LHT model \cite{Blanke:2009am} and 
RSc model \cite{Blanke:2008yr,Bauer:2009cf} with  respect to 
the SM but this is not guaranteed.
\end{itemize}

These patterns of flavour violations demonstrate very clearly the power of 
flavour physics in distinguishing different NP scenarios.
\boldmath
\subsection{EDMs, $(g-2)_\mu$ and $\mu\to e\gamma$}
\unboldmath
These three observables are governed by dipole operators but describe 
different physics as far as CP violation and flavour violation is concerned. 
EDMs are flavour conserving but CP-violating, $\mu\to e \gamma$ is CP-conserving but lepton flavour violating and finally $(g-2)_\mu$ is lepton flavour conserving and CP-conserving. A nice paper discussing all these observables 
simultaneously is \cite{Hisano:2009ae}.

In concrete models there exist correlations between these three observables 
of which EDMs and $\mu\to e\gamma$ are very strongly suppressed within the 
SM and have not been seen to date. $(g-2)_\mu$ on the other hand has been very precisely measured and exhibits a $3.2\sigma$ departure
 from the very precise SM value 
(see \cite{Prades:2009qp} and references therein).
Examples of these correlations can be found in 
\cite{Altmannshofer:2009ne,Altmannshofer:2008hc}. 
In certain supersymmetric 
flavour models with non-MFV interactions the solution of the $(g-2)_\mu$ 
anomaly implies simultaneously $d_e$ and $Br(\mu\to e \gamma)$ in the reach of experiments in this decade.

Here I would like only to report on correlations between $S_{\psi\phi}$ and the 
EDMs of the neutron, Thallium and Mercury atoms within the 
${\rm 2HDM_{\overline{MFV}}}$. 
The significant FBPhs required to reproduce the enhanced value of $S_{\psi\phi}$ 
in this model, necessarily  imply large EDMs in question. As a recent detailed
analysis in  \cite{Buras:2010zm} shows the present upper bounds on the
 EDMs do not forbid sizable non-standard CPV effects in $B_{s}$ mixing.
However, if a large CPV phase in $B_s$ mixing will be confirmed, this
will imply hadronic EDMs very close to their present experimental bounds,
within the reach of the next generation of experiments.
\subsection{News on right-handed currents}
One of the main properties of the Standard Model  regarding flavour 
 violating processes is the left-handed 
 structure of the charged currents that is 
 in accordance with the maximal violation 
 of parity observed in low energy processes. 
Yet, the SM is expected to be only the low-energy limit of a more fundamental
 theory in which  parity could be a good symmetry implying the existence of
 RH charged currents. Prominent examples of such fundamental 
 theories are  left-right symmetric models on which a rich literature exists.
We have also seen that several NP models that we discussed contain RH currents.

 The recent phenomenological interest in 
 the RH  currents in general, and not necessarily in the context of a
 given left-right 
 symmetric model as done recently in \cite{Maiezza:2010ic,Guadagnoli:2010sd},
 originated in tensions between inclusive and exclusive
 determinations of the elements of the CKM matrix $|V_{ub}|$ and  $|V_{cb}|$. 
 In particular it has been pointed out
 \cite{Crivellin:2009sd,Chen:2008se,Feger:2010qc}, 
 that the presence of RH  currents could either remove 
 or significantly weaken some of these tensions, especially in the 
 case of $|V_{ub}|$.

 Assuming that RH currents provide the solution to
 the problem at hand, there is an important question whether the strength
 of the RH currents required for this purpose is consistent with
 other flavour observables and whether it implies new effects somewhere else that
 could be used to test this idea more globally.

In order to answer this question an effective theory approach for the 
study of RH currents has been proposed in \cite{Buras:2010pz}.
In this approach 
 the central role is played by  a left-right symmetric flavour group
 $SU(3)_L \times SU(3)_R$, commuting with 
an underlying $SU(2)_L \times SU(2)_R \times U(1)_{B-L}$ global symmetry and
broken only by two Yukawa couplings. The model contains a new 
unitary matrix $V_R$  controlling flavour-mixing in the RH sector 
and can be considered as the minimally flavour violating generalization 
to the RH sector. Thus bearing in mind that this model contains non-MFV
interactions from the point of view of the standard MFV hypothesis that
includes only LH charged currents, we will call this model RHMFV.

A detailed analysis of this setup in \cite{Buras:2010pz} shows that the general structure of $V_R$ 
can be determined, under plausible assumptions, 
from the existing tree level decays in the $K$ and $B_d$ 
systems and FCNC processes. The presence of $(V-A)*(V+A)$ operators, 
whose contributions 
are strongly enhanced through renormalization group effects and in the case
of $\varepsilon_K$ also through chiral enhancement of their matrix elements, 
plays here an important role. The resulting $V_R$ differs significantly from 
the CKM matrix.

As already stated above the RHMFV model  goes beyond the MFV framework and new CPV phases in the 
RH sector allow for sizable enhancement of $S_{\psi\phi}$ and solution 
of the $\varepsilon_K$-anomaly as well as of the $|V_{ub}|$-problem. 
The resulting ``true'' value of $\sin 2\beta=0.77\pm0.05$ 
is much larger than 
the measured value of $S_{\psi K_S}=0.672\pm 0.023$. Usually this problem 
would be solved through a negative new phase $\varphi_{B_d}$, however the 
$\varepsilon_K$ constraint does not allow in this model 
 for a non-negligible value of 
this phase. It appears then that the simultaneous explanation of the 
$|V_{ub}|$-problem, of large $S_{\psi\phi}$ and of the data on $S_{\psi K_S}$ is 
problematic through RH currents alone.
Similarly in this simple setup the $B_{s,d}\to \mu^+\mu^-$  constraints 
eliminate the possibility of removing the known anomaly in 
$Z\to b \bar b$.

On top of it, the constraint from $B\to X_s l^+l^-$ precludes 
$B_{s}\to \mu^+\mu^-$ to be close to its present experimental bound.
 Moreover NP effects in $B_{d} \to \ell^+\ell^-$ are found generally 
smaller than in $B_{s} \to \ell^+\ell^-$.
Contributions from RH currents to 
$B \to \{X_s,K, K^*\} \nu\bar \nu$ and
$K\to\pi\nu\bar\nu$ decays can still be significant. 
Most important, the deviations from the SM in these decays 
would exhibit a well-defined pattern of correlations.
\boldmath
\subsection{Waiting for precise predictions of $\varepsilon'/\varepsilon$}
\unboldmath
The flavour studies of the last decade have shown that provided the hadronic 
matrix elements of QCD-penguin and electroweak penguin operators will be 
known with sufficient precision, $\varepsilon'/\varepsilon$ will play a very 
important role in constraining NP models. We have witnessed recently an 
impressive progress in the lattice evaluation of $\hat B_K$ that elevated 
$\varepsilon_K$ to the group of observables relevant for precision studies 
of flavour physics. Hopefully  this could also 
be the case of  $\varepsilon'/\varepsilon$ already in this decade.

\section{Summary}
We are at the beginning of a new decade which certainly will bring us 
first more detailed insights into the physics at short distance scales 
$10^{-19}-10^{-21}$m. The interplay of high energy collider results with 
the flavour precision experiments will allow us to make important steps 
towards a New Standard Model of which Flavour Theory will be a prominent 
part. For the time being we have to wait for the first big discoveries 
at the LHC and at other machines around the world. In particular we look 
forward to the 
full performance of the flavour superstars. These notes hopefully 
demonstrate that we will have a lot of fun with flavour physics in 
this decade.

\section*{Acknowledgements}
I would like to thank the organizers for inviting me to such a pleasant 
conference and all my collaborators for exciting time we spent together 
exploring the short distance scales with the help of flavour violating 
processes. In particular thanks go to Monika Blanke and Stefania Gori for
reading carefully the manuscript of this paper.
This research was partially supported by the Cluster of Excellence `Origin and Structure
of the Universe' and  by the German `Bundesministerium f\"ur Bildung und Forschung'
under contract 05H09WOE.


\begin{footnotesize}

\end{footnotesize}



\begin{thebibliography}{99}



\bibitem{Buras:2009if}
  A.~J.~Buras,
  PoS E {\bf PS-HEP2009} (2009) 024
  [arXiv:0910.1032 [hep-ph]].

\bibitem{Isidori:2010kg}
  G.~Isidori, Y.~Nir and G.~Perez,
  arXiv:1002.0900 [hep-ph];  
        O.~Gedalia and G.~Perez,
        arXiv:1005.3106 [hep-ph].


\bibitem{Cabibbo:1963yz}
  N.~Cabibbo,
  Phys.\ Rev.\ Lett.\  {\bf 10} (1963) 531.
  M.~Kobayashi and T.~Maskawa,
  Prog.\ Theor.\ Phys.\  {\bf 49} (1973) 652.



\bibitem{Glashow:1970gm}
  S.~L.~Glashow, J.~Iliopoulos and L.~Maiani,
  Phys.\ Rev.\  D {\bf 2} (1970) 1285.


\bibitem{Fleischer:2007wg}
  R.~Fleischer and M.~Gronau,
  Phys.\ Lett.\  B {\bf 660} (2008) 212
  [arXiv:0709.4013 [hep-ph]].


\bibitem{Lenz:2010gu}
  A.~Lenz {\it et al.},
  arXiv:1008.1593 [hep-ph]; 
{\sf ckmfitter.in2p3.fr}



\bibitem{D'Ambrosio:2002ex}
  G.~D'Ambrosio, G.~F.~Giudice, G.~Isidori and A.~Strumia,
  Nucl.\ Phys.\  B {\bf 645} (2002) 155
  [arXiv:hep-ph/0207036].


\bibitem{Buras:2000dm}
  A.~J.~Buras, P.~Gambino, M.~Gorbahn, S.~Jager and L.~Silvestrini,
  Phys.\ Lett.\  B {\bf 500} (2001) 161
  [arXiv:hep-ph/0007085].
  A.~J.~Buras,
  Acta Phys.\ Polon.\  B {\bf 34} (2003) 5615
  [arXiv:hep-ph/0310208].



\bibitem{Kagan:2009bn}
  A.~L.~Kagan, G.~Perez, T.~Volansky and J.~Zupan,
  Phys.\ Rev.\  D {\bf 80} (2009) 076002
  [arXiv:0903.1794 [hep-ph]].

\bibitem{Colangelo:2008qp}
  G.~Colangelo, E.~Nikolidakis and C.~Smith,
  Eur.\ Phys.\ J.\  C {\bf 59} (2009) 75
  [arXiv:0807.0801 [hep-ph]].



\bibitem{Mercolli:2009ns}
  L.~Mercolli and C.~Smith,
  Nucl.\ Phys.\  B {\bf 817} (2009) 1
  [arXiv:0902.1949 [hep-ph]].

\bibitem{Paradisi:2009ey}
  P.~Paradisi and D.~M.~Straub,
  Phys.\ Lett.\  B {\bf 684} (2010) 147
  [arXiv:0906.4551 [hep-ph]].


\bibitem{Ellis:2007kb}
  J.~R.~Ellis, J.~S.~Lee and A.~Pilaftsis,
  Phys.\ Rev.\  D {\bf 76} (2007) 115011
  [arXiv:0708.2079 [hep-ph]].


\bibitem{Buras:2010mh}
  A.~J.~Buras, M.~V.~Carlucci, S.~Gori and G.~Isidori,
  arXiv:1005.5310 [hep-ph].

\bibitem{Aaltonen:2007he}
  T.~Aaltonen {\it et al.}  [CDF Collaboration],
  Phys.\ Rev.\ Lett.\  {\bf 100} (2008) 161802
  [arXiv:0712.2397 [hep-ex]].
  V.~M.~Abazov {\it et al.}  [D0 Collaboration],
  Phys.\ Rev.\ Lett.\  {\bf 101} (2008) 241801
  [arXiv:0802.2255 [hep-ex]].



\bibitem{Abazov:2010hv}
  V.~M.~Abazov {\it et al.}  [D0 Collaboration],
  Phys.\ Rev.\  D {\bf 82} (2010) 032001
  [arXiv:1005.2757 [hep-ex]].
  V.~M.~Abazov {\it et al.}  [D0 Collaboration],
  Phys.\ Rev.\ Lett.\  {\bf 105} (2010) 081801
  [arXiv:1007.0395 [hep-ex]].


\bibitem{Aaltonen:2010xx}
T.~Aaltonen {\it et al.}  [CDF Collaboration], CDF public notes, 9458, 10206.
  V.~M.~Abazov {\it et al.}  [D0 Collaboration], D0 Conference note 6098.
 
\bibitem{Lunghi:2008aa}
  E.~Lunghi and A.~Soni,
  Phys.\ Lett.\  B {\bf 666} (2008) 162
  [arXiv:0803.4340 [hep-ph]].

\bibitem{Buras:2008nn}
  A.~J.~Buras and D.~Guadagnoli,
  Phys.\ Rev.\  D {\bf 78}, 033005 (2008)
  [arXiv:0805.3887 [hep-ph]];
  Phys.\ Rev.\  D {\bf 79} (2009) 053010
  [arXiv:0901.2056 [hep-ph]].






\bibitem{Batell:2010qw}
  B.~Batell and M.~Pospelov,
  arXiv:1006.2127 [hep-ph].





\bibitem{Buras:2010zm}
  A.~J.~Buras, G.~Isidori and P.~Paradisi,
  arXiv:1007.5291 [hep-ph].

\bibitem{Hou:2005yb}
  W.~S.~Hou, M.~Nagashima and A.~Soddu,
  Phys.\ Rev.\  D {\bf 72} (2005) 115007
  [arXiv:hep-ph/0508237].
  Phys.\ Rev.\  D {\bf 76} (2007) 016004
  [arXiv:hep-ph/0610385].


\bibitem{Soni:2008bc}
  A.~Soni, A.~K.~Alok, A.~Giri, R.~Mohanta and S.~Nandi,
  Phys.\ Lett.\  B {\bf 683} (2010) 302
  [arXiv:0807.1971 [hep-ph]].
        A.~Soni, A.~K.~Alok, A.~Giri, R.~Mohanta and S.~Nandi,
        arXiv:1002.0595 [hep-ph].



      \bibitem{Bobrowski:2009ng}
        M.~Bobrowski, A.~Lenz, J.~Riedl and J.~Rohrwild,
        Phys.\ Rev.\  D {\bf 79} (2009) 113006
        [arXiv:0902.4883 [hep-ph]].
  O.~Eberhardt, A.~Lenz and J.~Rohrwild,
  arXiv:1005.3505 [hep-ph].


\bibitem{Buras:2010pi}
  A.~J.~Buras, B.~Duling, T.~Feldmann, T.~Heidsieck, C.~Promberger and S.~Recksiegel,
  arXiv:1002.2126 [hep-ph];
  JHEP {\bf 1007} (2010) 094
  [arXiv:1004.4565 [hep-ph]].
        A.~J.~Buras, B.~Duling, T.~Feldmann, T.~Heidsieck and C.~Promberger,
        arXiv:1006.5356 [hep-ph].






   \bibitem{Buras:2010pz}
        A.~J.~Buras, K.~Gemmler and G.~Isidori,
        arXiv:1007.1993 [hep-ph].



\bibitem{Maiezza:2010ic}
  A.~Maiezza, M.~Nemevsek, F.~Nesti and G.~Senjanovic,
  arXiv:1005.5160 [hep-ph].



\bibitem{Guadagnoli:2010sd}
  D.~Guadagnoli and R.~N.~Mohapatra,
  arXiv:1008.1074 [hep-ph].


\bibitem{Crivellin:2009sd}
  A.~Crivellin,
  Phys.\ Rev.\  D {\bf 81} (2010) 031301
  [arXiv:0907.2461 [hep-ph]].

\bibitem{Chen:2008se}
  C.~H.~Chen and S.~h.~Nam,
  Phys.\ Lett.\  B {\bf 666} (2008) 462
  [arXiv:0807.0896 [hep-ph]].

\bibitem{Feger:2010qc}
  R.~Feger, V.~Klose, H.~Lacker, T.~Lueck and T.~Mannel,
  arXiv:1003.4022 [hep-ph].

\bibitem{Altmannshofer:2009ne}
  W.~Altmannshofer, A.~J.~Buras, S.~Gori, P.~Paradisi and D.~M.~Straub,
  Nucl.\ Phys.\  B {\bf 830} (2010) 17
  [arXiv:0909.1333 [hep-ph]].

\bibitem{Lalak:2010bk}
  Z.~Lalak, S.~Pokorski and G.~G.~Ross,
  arXiv:1006.2375 [hep-ph].

\bibitem{Dudas:2010yh}
  E.~Dudas, G.~von Gersdorff, J.~Parmentier and S.~Pokorski,
  arXiv:1007.5208 [hep-ph].




\bibitem{Glashow:1976nt}
  S.~L.~Glashow and S.~Weinberg,
  Phys.\ Rev.\  D {\bf 15} (1977) 1958.

\bibitem{Paschos:1976ay}
  E.~A.~Paschos,
  Phys.\ Rev.\  D {\bf 15} (1977) 1966.

\bibitem{Botella:2009pq}
  F.~J.~Botella, G.~C.~Branco and M.~N.~Rebelo,
  Phys.\ Lett.\  B {\bf 687} (2010) 194
  [arXiv:0911.1753 [hep-ph]].




\bibitem{Pich1}
  A.~Pich and P.~Tuzon,
  Phys.\ Rev.\  D {\bf 80} (2009) 091702
  [arXiv:0908.1554 [hep-ph]].
  M.~Jung, A.~Pich and P.~Tuzon,
  arXiv:1006.0470 [hep-ph].




\bibitem{Dobrescu:2010rh}
  B.~A.~Dobrescu, P.~J.~Fox and A.~Martin,
  Phys.\ Rev.\ Lett.\  {\bf 105} (2010) 041801
  [arXiv:1005.4238 [hep-ph]].

\bibitem{Braeuninger:2010td}
  C.~B.~Braeuninger, A.~Ibarra and C.~Simonetto,
  Phys.\ Lett.\  B {\bf 692} (2010) 189
  [arXiv:1005.5706 [hep-ph]].



\bibitem{Buras:1998raa}
  A.~J.~Buras,
  arXiv:hep-ph/9806471.


\bibitem{Brod:2010mj}
  J.~Brod and M.~Gorbahn,
  arXiv:1007.0684 [hep-ph].


\bibitem{Brod:2010hi}
  J.~Brod, M.~Gorbahn and E.~Stamou,
  arXiv:1009.0947 [hep-ph].



\bibitem{Antonio:2007pb}
  D.~J.~Antonio {\it et al.}  [RBC Collaboration and UKQCD Collaboration],
  Phys.\ Rev.\ Lett.\  {\bf 100} (2008) 032001
  [arXiv:hep-ph/0702042].
  C.~Aubin, J.~Laiho and R.~S.~Van de Water,
  Phys.\ Rev.\  D {\bf 81} (2010) 014507
  [arXiv:0905.3947 [hep-lat]].
  T.~Bae {\it et al.},
  arXiv:1008.5179 [hep-lat].





\bibitem{Grzadkowski:2010es}
  B.~Grzadkowski, M.~Iskrzynski, M.~Misiak and J.~Rosiek,
  arXiv:1008.4884 [hep-ph].




\bibitem{UTfit-web}
  See the talk by Cecilia Tarantino at ICHEP 2019 and {\sf www.utfit.org}





\bibitem{Buras:2010pza}
  A.~J.~Buras, D.~Guadagnoli and G.~Isidori,
  Phys.\ Lett.\  B {\bf 688} (2010) 309
  [arXiv:1002.3612 [hep-ph]].



\bibitem{Ball:2006xx}
  P.~Ball and R.~Fleischer,
  Eur.\ Phys.\ J.\  C {\bf 48} (2006) 413
  [arXiv:hep-ph/0604249].



 



\bibitem{Ligeti:2010ia}
  Z.~Ligeti, M.~Papucci, G.~Perez and J.~Zupan,
  arXiv:1006.0432 [hep-ph].


\bibitem{Blum:2010mj}
  K.~Blum, Y.~Hochberg and Y.~Nir,
  arXiv:1007.1872 [hep-ph].


\bibitem{Bauer:2010dg}
  C.~W.~Bauer and N.~D.~Dunn,
  arXiv:1006.1629 [hep-ph].
  N.~G.~Deshpande, X.~G.~He and G.~Valencia,
  arXiv:1006.1682 [hep-ph].
  D.~Choudhury and D.~K.~Ghosh,
  arXiv:1006.2171 [hep-ph].
  C.~H.~Chen, C.~Q.~Geng and W.~Wang,
  arXiv:1006.5216 [hep-ph].
  P.~Ko and J.~h.~Park,
  arXiv:1006.5821 [hep-ph].
  S.~F.~King,
  arXiv:1006.5895 [hep-ph].
  Y.~Bai and A.~E.~Nelson,
  arXiv:1007.0596 [hep-ph].
  J.~Kubo and A.~Lenz,
  arXiv:1007.0680 [hep-ph].
  C.~Berger and L.~M.~Sehgal,
  arXiv:1007.2996 [hep-ph].
  B.~Dutta, Y.~Mimura and Y.~Santoso,
  arXiv:1007.3696 [hep-ph].
  S.~Oh and J.~Tandean,
  arXiv:1008.2153 [hep-ph].







 


\bibitem{Blanke:2009am}
  M.~Blanke, A.~J.~Buras, B.~Duling, S.~Recksiegel and C.~Tarantino,
  Acta Phys.\ Polon.\  B {\bf 41} (2010) 657
  [arXiv:0906.5454 [hep-ph]].

\bibitem{Blanke:2008yr}
  M.~Blanke, A.~J.~Buras, B.~Duling, K.~Gemmler and S.~Gori,
  JHEP {\bf 0903} (2009) 108
  [arXiv:0812.3803 [hep-ph]].
  M.~Blanke, A.~J.~Buras, B.~Duling, S.~Gori and A.~Weiler,
  JHEP {\bf 0903} (2009) 001
  [arXiv:0809.1073 [hep-ph]].

\bibitem{Bauer:2009cf}
  M.~Bauer, S.~Casagrande, U.~Haisch and M.~Neubert,
  arXiv:0912.1625 [hep-ph].

 
  \bibitem{Hisano:2009ae}
  J.~Hisano, M.~Nagai, P.~Paradisi and Y.~Shimizu,
  JHEP {\bf 0912} (2009) 030
  [arXiv:0904.2080 [hep-ph]].



\bibitem{Prades:2009qp}
  J.~Prades,
  Acta Phys.\ Polon.\ Supp.\  {\bf 3} (2010) 75
  [arXiv:0909.2546 [hep-ph]].




\bibitem{Altmannshofer:2008hc}
  W.~Altmannshofer, A.~J.~Buras and P.~Paradisi,
  Phys.\ Lett.\  B {\bf 669} (2008) 239
  [arXiv:0808.0707 [hep-ph]].



\end{thebibliography}
\end{document}